\documentstyle[preprint,aps]{revtex}

\begin{document}
\title{On the ground state of solids with strong electron correlations}

\author{P. Fulde, H. Stoll* and K. Kladko**}

\address{
Max-Planck-Institut f\"ur Physik komplexer Systeme,
N\"othnitzer Stra\ss e 38,
01187 Dresden (Germany)\\
Institut f\"ur Theoretische Chemie der Universit\"at Stuttgart,
70550 Stuttgart (Germany)*\\
Los Alamos National Laboratory,
Los Alamos, NM 87545 (USA)**}

\date{\today}
\maketitle

\abstract

We formulate the calculation of the ground-state wavefunction and energy of a
system of strongly correlated electrons in terms of scattering matrices. A
hierarchy of approximations is introduced which results in an incremental
expansion of the energy. The present approach generalizes previous work
designed for weakly correlated electronic systems.

\newpage

\newcommand{\gsim}{\mathrel{\raise.3ex\hbox{$>$\kern-.75em\lower1ex\hbox{$\sim$}}}}
\newcommand{\lsim}{\mathrel{\raise.3ex\hbox{$<$\kern-.75em\lower1ex\hbox{$\sim$}}}}

{\section{\bf Introduction}}

\vspace{1cm}

A microscopic wavefunction-based description of electron correlations in the
ground state of extended systems, i.e., large molecules or solids remains a
challenging problem. This holds particularly true when the correlations are
strong. Much progress has been achieved though over the last twenty years. For
example, it has become clear that local operators have to be applied for
describing the correlation hole of the electrons [1-6]. Otherwise calculations
with controlled approximations for extended systems become nonfeasible. When
starting from a self-consistent field (SCF) wavefunction for the ground state
denoted by $\mid\Phi_0>$ the wave operator $\tilde{\Omega}$, which transforms
it into the true ground-state wavefunction $\mid\Psi_0>$ must be constructed
from $local$ operators $A_{\nu}$. They describe one-, two- or generally
multiparticle excitations out of $\mid\Phi_0>$. The fact that the $A_{\nu}$ are
local operators implies that the creation and annihilation operators appearing
in $A_{\nu}$ refer to $local$ orbitals instead of $canonical$ orbitals or Bloch
states. 

The first ground-state calculations based on the use of local operators were
done for diamond [7]. However, it was not until the method of increments was
pointed out [5,8], that a large number of solids were successfully treated by
quantum-chemical configuration-interaction (CI) techniques (see, e.g., [8-10]). Although
most of the calculations were done with respect to the ground-state
wavefunction it was shown that the same ideas can be also applied to the
calculation of excited states, i.e., energy bands of a solid [11]. 

The calculations described above were done for systems in which electron
correlations are not too strong, i.e., for which the SCF configuration is a
good starting-point. If the electronic correlations are strong one should not
start from the independent-electron approximation when attempting to calculate
the true ground-state wavefunction. It has been suggested that the method
of increments can be also applied to strongly correlated systems by performing
multi-configuration SCF (MCSCF) or complete-active-space SCF (CASSCF)
calculations for localized orbital groups [12]. But the theoretical formulation
of such an approximation scheme as well as the form of the ground-state wave
function have remained unclear. 

The aim of the present communication is to provide a basis for the incremental
method when the correlations are strong. For that purpose we have to modify a
derivation of an incremental scheme given in Ref. [13]. We want to show how one
can construct a cumulant wave operator for strongly correlated electron
systems. This operator defines the ground-state of the system.

The paper is organized as follows. In the next section a frame is provided for
the computation of the ground state of a strongly correlated extended electron
system by means of quantum-chemical methods. Section III demonstrates
explicitly how the calculations have to be done in practice. Section IV
contains a summary and the conclusions.

\vspace{1cm}

{\section{\bf Formalism}}

\vspace{0.5cm} 

Starting point is the Hamiltonian $H$ of the electronic system

\begin{equation}
H~=~\sum_{ij\sigma} t_{ij}a^+_{i\sigma}a_{j\sigma}~+~\frac{1}{2}\sum_{ijk\ell
\sigma\sigma'} V_{ijk\ell}a^+_{i\sigma}
a^+_{k\sigma'}a_{\ell\sigma'}a_{j\sigma}.  
\end{equation}

It refers to a given basis set $f_i(\underline{r})$ for which usually
Gauss-type (GTO) or Slater-type orbitals are chosen. The matrix elements
$t_{ij}$ and $V_{ijk\ell}$ refer to the one-electron (including kinetic energy)
and two-electron interaction energy, respectively. We split $H$ into a
self-consistent field part $H_{SCF}$ and a residual interaction part
$H_{res}$. The latter is given, for the closed-shell case, e.g., by 

\begin{eqnarray}
H_{res} & = & \sum_{ijk\ell}~ \left[ ~\frac{1}{2}\sum_{\sigma\sigma'}
V_{ijk\ell}a^+_{i\sigma}a^+_{k\sigma'}a_{\ell\sigma'}a_{j\sigma}-\sum_{\sigma}
\left( V_{ijk\ell}-\frac{1}{2} V_{i\ell kj} \right)
P_{k\ell}a_{i\sigma}^+a_{j\sigma} \right.\\
\nonumber 
& + & \left. \frac{1}{2}~ \left( V_{ijk\ell}-\frac{1}{2}V_{i\ell kj} \right)
P_{ij}P_{k\ell}~ \right].
\end{eqnarray}

The one-particle density matrix $P_{ij}$ is defined by

\begin{equation}
P_{ij}~=~\sum_{\sigma}<a^+_{i\sigma}a_{j\sigma}>
\end{equation}

and $<...>=<\Phi_{SCF}\mid ...\mid\Phi_{SCF}>$ with $\mid\Phi_{SCF}>$ denoting
the ground state of $H_{SCF}$. It is noticed that $H_{res}$ consists of a
constant term plus one- and two-particle excitations. We subdivide $H_{res}$
further into 

\begin{equation}
H_{res}=\sum_I~H_I+\sum_{I, J}~H_{IJ}.
\end{equation}

Here $H_I$ denotes that part of $H_{res}$ where excitations are restricted to a
group of localized orbitals belonging to center $I$ (e.g., an atom or a bond),
while $H_{IJ}$ contains the part in which the centers $I$ and $J$ are involved
in the process of creating holes. For example, van der Waals type interactions
between atoms (or bonds) $I$ and $J$ belong to $H_{IJ}$ since they involve a
one-particle excitation on each of the atoms (bonds) $I$ and $J$. This division
of $H_{res}$ differs from the one in Ref. [13] where $H_{res}$ was decomposed
into contributions corresponding to different pairs of holes generated out of
the self-consistent ground state $\mid\Phi_{SCF}>$.

We characterize the ground state of $H$ by the cumulant wave operator
$\mid\Omega)$ [14, 6]. The conventional wave operator $\tilde{\Omega}$ is
defined as the one which transforms $\mid\Phi_{SCF}>$ into the exact ground
state $\mid\Psi_0>$, i.e., $\mid\Psi_0>=\tilde{\Omega}\mid\Phi_{SCF}>$. The
cumulant wave operator $\mid\Omega)$ differs from $\tilde{\Omega}$ in that it
is defined in a space with a different metric than that, e.g., of the
conventional Hilbert or operator space. Here the metric is defined by the
following bilinear form of two general operators $A$ and $B$. 

\begin{equation}
(A\mid B)~=~<\Phi_{SCF}\mid A^+B\mid\Phi_{SCF}>^c.
\end{equation}

The upper script $c$ implies taking the cumulant of that expression [14,15,6].
The usefulness of cumulants in quantum mechanics was stressed by Kubo [15] who
generalized earlier work of Ursell and Mayer in classical statistical mechanics
[16]. Cumulants have the advantage that they correspond to linked
clusters. Therefore the problem of size consistency (or extensivity) does not
exist when quantities are expressed in terms of them. For further information
we refer to, e.g., Ref.[6]. In accordance with the above the ground-state
energy is given by 

\begin{equation}
E_0~=~(H\mid\Omega).
\end{equation}

The cumulant wave operator is of the form [6]

\begin{equation}
\mid\Omega)~=~ \left. \lim_{z\to 0} \left| 1+\frac{1}{z-H}~ \right. H_{res}
\right) . 
\end{equation}

In accordance with Ref. [13] we define a scattering operator

\begin{equation}
\mid S)~=~\mid\Omega-1)~=~\lim_{z\to
0}\sum_{n=1}^{\infty} \left. \left| \left( \frac{1}{z-H_{SCF}}H_{res} \right)
^n \right. \right) . 
\end{equation}

In proceeding we use an argument developed in [13] which can be considered as a
generalization of Faddeev's equation. Faddeev derived an equation [17] which
expresses the scattering operator of a three-particle system in terms of the
scattering matrices for the different two-particle channels. For the latter
often analytic solutions can be found. More generally, we aim at expressing the
scattering operator of an $\tilde{N}$ electron system in terms of scattering
matrices of simpler subsystems. With this goal in mind we introduce the
operators 

\begin{equation}
A_{II}~=~\lim_{z\to 0}\frac{1}{z-H_{SCF}} H_I~,~A_{IJ}~=~\lim_{z\to
0}\frac{1}{z-H_{SCF}}H_{IJ}. 
\end{equation}

With their help we rewrite

\begin{equation}
\mid S)~=~\sum_{n=1}^{\infty}~ 
\left.
\left| \left( 
\sum^N_{I,J} A_{IJ}
\right) ^n \right. 
\right) 
\end{equation}

where $N$ is the number of centers, i.e., atoms (bonds) in the system. We
introduce Greek letters to denote  pair labels $IJ$ so that
$\sum\limits_{IJ}\mid A_{IJ}> = \sum\limits_\alpha \mid A_\alpha>$ and
decompose $\mid S )$ into terms with one Greek index and a remaining
part. Therefore we write 
 
\begin{eqnarray}
S & = & \sum_\alpha~ \left( \sum^{\infty}_{n=1}A^n_\alpha \right) +
\sum_{\alpha \neq \beta}T_{\alpha \beta}\\ \nonumber
~ & = & \sum_\alpha S_\alpha + \sum_{\alpha \neq \beta} T_{\alpha \beta}.
\end{eqnarray}

The $T_{\alpha \beta}$ are defined as follows: When (10) is decomposed into
different terms we include in $T_{\alpha \beta}$ all those which begin with
$A_\alpha$ from the left followed by $A_\beta$ as the first factor different
from $A_\alpha$. For example, terms of the form $A_\alpha A_\beta ..., A_\alpha
A_\alpha A_\beta ..., A_\alpha A_\beta A_\beta ...$ are all included in
$T_{\alpha \beta}$. With this definition we can write

\begin{eqnarray}
T_{\alpha \beta} & = & (A_\alpha A_\beta + A_\alpha A_\beta A_\alpha +
...) \cdot\\ \nonumber
~ & \cdot & \left( 1 + \sum_{\gamma \neq \alpha,\beta} S_\gamma + \sum_{\gamma
\neq \alpha,\beta; \delta} T_{\gamma \delta} \right) .
\end{eqnarray}

When adding $T_{\alpha \beta} + T_{\beta \alpha}$ one notices that the first
bracket is nothing else but the scattering operator $S_{\alpha \beta}$ of a
Hamiltonian $H_{SCF} + H_\alpha + H_\beta$, except that the contributions
$S_\alpha + S_\beta$ are missing. Therefore we can write

\begin{eqnarray*}
T_{\alpha \beta} + T_{\beta \alpha}~=~(S_{\alpha \beta} - S_\alpha - S_\beta)
\left( 1 + \sum_{\gamma \neq \alpha,\beta} S_\gamma + \sum_{\gamma \neq
\alpha,\beta; \delta} T_{\gamma \delta} \right) .\\ 
\end{eqnarray*}

or

\begin{equation}
S~ =~ \sum_\alpha S_\alpha~+~\sum_{<\alpha \beta>}~(S_{\alpha \beta} -
S_\alpha - S_\beta) \left( 1~+~\sum_{\gamma \neq \alpha, \beta}S_\gamma +
\sum_{\gamma \neq \alpha,\beta; \delta}T_{\gamma \delta} \right) . 
\end{equation}

Here $<\alpha \beta>$ denotes different pairs.

One expects that the matrix elements of $H_I$ are generally much larger than
those of $H_{IJ}$. Therefore it seems advantageous to introduce
approximations to $\mid S)$ by resumming the right hand side of the last
equation according to different numbers of sites involved. To lowest order
$\mid S)$ is therefore given by

\begin{equation}
\mid S) = \sum_I S_I,
\end{equation}

where $S_I$ ($= S_\alpha$ with $\alpha = II$) is the scattering operator of
Hamiltonian $H_{SCF} + H_I$. 

In this single-site approximation

\begin{eqnarray}
E_0 & = & E_{SCF} + (H \mid S)\\ \nonumber
~   & = & E_{SCF} + \sum_I \epsilon_I
\end{eqnarray}

with $\epsilon_I = (H \mid S_I)$. In next order we include the terms $S_\alpha$
with $\alpha = IJ$ and $T_{\alpha \beta}$ with $\alpha$ and $\beta$ being $II,
JJ, IJ$ and $JI$. In this two-sites approximation the second bracket in (12) is
replaced by unity. By adding up the different contributions we find

\begin{eqnarray*}
\mid S)~=~\sum_I~\mid S_I) + \sum_{<IJ>} \mid S_{IJ} - S_I - S_J)
\end{eqnarray*}

where $S_{IJ}$ is the scattering operator belonging to $H_{SCF} + H_I + H_J +
H_{IJ} + H_{JI}$.

Furthermore,

\begin{equation}
E_0 = E_{SCF} + \sum_I \epsilon_I + \sum_{<IJ>} \epsilon_{IJ}
\end{equation}

with $\epsilon_{IJ} = (H \mid S_{IJ}) - \epsilon_I - \epsilon_J$. This
procedure can be continued. By including also three-sites terms we find

\begin{equation}
\mid S)~=~\sum_I~\mid S_I)+\sum_{<IJ>}\mid\delta
S_{IJ})+\sum_{<IJK>}~\mid\delta S_{IJK})+...\\ 
\end{equation}

with $\mid\delta S_{IJ})=\mid S_{IJ})-\mid S_I)-\mid S_J)$ as before and

\begin{equation}
\mid\delta S_{IJK})=\mid S_{IJK})-\mid\delta S_{IJ})-\mid\delta
S_{JK})-\mid\delta S_{KI})-\mid S_I)-\mid S_J)-\mid S_K). 
\end{equation} 

Here $\mid S_{IJK})$ is the scattering operator of a Hamiltonian
$H=H_{SCF}+H_I+H_J+H_K+H_{IJ}+H_{IK}+H_{JK}+H_{JI}+H_{KI}+H_{KJ} $.
This approximation scheme can be continued until finally the exact scattering
operator $\mid S)$ is obtained. The corresponding energy expression (15) is
then 

\begin{equation}
E_0~=~E_{SCF}+\sum_I\epsilon_I+\sum_{<IJ>}\epsilon_{IJ}+\sum_{<IJK>}\epsilon_{IJK}+...
\end{equation}

with

\begin{equation}
\epsilon_{IJK}=(H\mid S_{IJK})-\epsilon_{IJ}-\epsilon_{IK}-\epsilon_{JK}-\epsilon_I-\epsilon_J-\epsilon_K
\end{equation}

etc.\\

The advantage of the above formalism is that we have reduced the ground-state
calculations for extended systems like solids to the computation of
single-center, two-center etc. scattering matrices. These matrices can be
determined by means of quantum-chemical methods whereby all the other electrons
in $\mid \Phi_{SCF}>$ are kept frozen. There is no difficulty in calculating the
scattering matrices also when the electrons are strongly correlated. For
example, multiconfiguration SCF (MCSCF) calculations or complete-active-space
SCF (CASSCF) calculations, followed by a multireference CI (MRCI) treatment,
serve that purpose. Strong correlations can therefore be treated with a high
degree of accuracy in ground-state calculations for solids.

\vspace{1cm}

{\section{\bf Applications}}

\vspace{0.5cm} 

In the following we want to outline in some more detail how ground-state
calculations for strongly correlated electron systems have to be performed. We
limit ourselves to insulators or semiconductors, i.e., systems with a gap in
the excitation spectrum. Starting point is the Hamiltonian (1) acting in a
space spanned by a properly chosen basis set of GTO's. 
After a SCF calculation has been performed, e.g., by using the program package
CRYSTAL [18] or the code developed by Shukla et al.\ [19], one has to express
the SCF orbitals in the form of orthogonal localized Wannier orbitals. This is
achieved either by an a-posteriori localization procedure if CRYSTAL is used,
or by using Shukla's program which yields directly the occupied SCF orbitals in
localized form. The corresponding creation and annihilation operators are
denoted by $\tilde{c}^+_{\ell\sigma},\tilde{c}_{\ell\sigma}$. For the strongly
correlated electrons in the system, e.g., the $d$ electrons of a transition
metal atom or the $f$ electrons of a rare earth or actinide atom, we need to
express in localized form not only the occupied, but also the unoccupied
(virtual) $d$ or $f$ orbitals. Finding them poses no problem and we include
them in the set of operators
$\tilde{c}^+_{\ell\sigma},\tilde{c}_{\ell\sigma}$. In a next step the residual
interactions $H_{res}$ are expressed in terms of the
$\tilde{c}_{\ell\sigma}^+,\tilde{c}_{\ell\sigma}$ operators.
Except for the strongly correlated electrons we express only the annihilation
operators in (2) in terms of the $\tilde{c}_{\ell\sigma}$.
(The external orbitals have to be localized, but they need not being orthogonal
to each other, cf.\ e.g.\ [20].) The strong correlations are treated by a
CASSCF calculation. The choice of the active space depends on the scattering
matrix we want to calculate. For the determination of the $S_I$ we choose the
strongly correlated orbitals of that center (e.g., the localized $d$ or $f$
orbitals of an atom or the bonding and anti-bonding orbitals of a bond) for the
active space. With all electrons kept frozen except those on center $I$, the CASSCF
calculation with the active space involving orbitals on that center only
accounts for the strong intra-atomic (or intra-bond) correlations. When
$S_{IJ}$ is calculated we must include in the active space the localized $d(f)$
orbitals (or bonding/antibonding orbitals) on centers $I$ $and$ $J$. The
resulting ground state is denoted by $\mid\Psi_0>$. After every CASSCF
calculation, the remaining (weak) correlations may be accurately taken into
account by means of calculations within some variant of the coupled electron
pair approximation (e.g., MRCEPA or MRACPF with single and double excitations into the external space).
In case that the two-center scattering matrices $S_{IJ}$ are not sufficient for
an accurate determination of $\mid\Omega)$ one can extend the calculations to
three-center scattering matrices $S_{IJK}$. 

Let us apply now the above considerations to simple molecular examples
involving hydrocarbons. We address the question of how to properly treat such
systems in the strong-correlation limit of dissociation into separate atoms;
the ultimate goal of our investigation being a unified treatment of C$_\infty$
(diamond) for a wide range of C-C internuclear distances. Specifically, we
performed calculations for CH$_4$, C$_2$H$_6$, C$_3$H$_8$, and C$_5$H$_{12}$
(neopentane); all angles were kept fixed at 109.47$^0$; standard equilibrium
C-H and C-C bond lengths of 1.102 and 1.544 \AA\ , respectively, were uniformly
scaled by factors $f$, with $1 \leq f \leq 100$.  We concentrated on strong
correlations, i.e., restricted our basis set to the single-zeta level
((9s4p)/[2s1p] for C, (4s)/[1s] for H, using subsets of Dunning's correlation
consistent valence double-zeta sets [21]). Since SCF calculations, even with
such small basis sets, encounter severe convergence difficulties for large
internuclear distances (and yield physically unreasonable energies anyway, high
above the dissociation limit), we started from localized two-center orbitals
generated as follows. We combined $sp^3$ hybrids on the C centers with each
other and with H $1s$ orbitals to form bonding and anti-bonding LMOs
(coefficients $\pm 1$) between next-neighbour atoms; we then
Gram-Schmidt-orthogonalized all valence orbitals to the C $1s$ cores,
symmetrically orthogonalized the bonding LMOs among each other, proceeded by
Gram-Schmidt-orthogonalizing the anti-bonding LMOs to the bonding ones, and
finally symmetrically orthogonalized within the anti-bonding space. With this
construction, we can build up an SCF-like closed-shell state, with all bonding
LMOs doubly occupied, which should resemble the true SCF ground state of
C$_\infty$, in the limit of large cluster size (and in a minimal-basis set
representation, of course). We then defined groups of orbitals, pairing each of
the bonding LMOs with the corresponding anti-bonding one, and performed CASCI
calculations [22 - 24] with one of the groups active in turn --- this leads to
correlation-energy increments $\Delta\epsilon_{CC}$ and $\Delta\epsilon_{CH}$
describing the breaking of a CC or a CH bond in a frozen closed-shell
environment. Their sum provides us with a first approximation to the
correlation energy of the system, but still not a very good one, since
reorganization at the C atoms (leading to $^3$P ground states in the limit of
$R \rightarrow \infty$) is not taken into account. We therefore introduced, in
the next step, an atomic correction by correlating simultaneously all eight
LMOs (bonding and anti-bonding ones) related to a given C atom, in a CASCI
calculation. This defines an atomic increment $\Delta\Delta\epsilon_C =
\epsilon_C - \sum_i \Delta\epsilon_{CX}$, where $\epsilon_C$ is the correlation
energy of the calculation just mentioned, and the $\Delta\epsilon_{CX}$ are the
single-bond increments of the neighbouring atoms X = C, H. Again, by adding up
all $\Delta\Delta\epsilon_C$ contributions, we obtain an improved estimate for
the energy of the system. (The next step -- which we did not perform any more --
would be to determine non-additivity corrections for pairs, triples etc.\ of atomic
increments. In the limit of $n$-tuple corrections ($n \rightarrow \infty$), this
should lead to the full-CI energy of the system, irrespective of the starting-point
chosen, i.e., irrespective of the fact that we did not start from a variational
SCF wavefunction.) 

In Table 1, the so-obtained correlation-energy estimates for CH$_4$ and
C$_2$H$_6$ are compared to reference values (full valence CI results). The
errors of the 'SCF' energies are huge, of course ($\sim$ 1.5 a.u. for $f=100$
and $\sim$ 0.13 a.u. for $f=1$, in the case of CH$_4$); about half of the error
for $f=1$ is due to the non-self-consistent preparation of the ground state.
Including the bond correlations, $\Delta\epsilon_{CC}$ and
$\Delta\epsilon_{CH}$, errors are reduced by nearly an order of magnitude (to
$\sim$ 0.25 a.u. for $f=100$ and $\sim$ 0.02 a.u. for $f=1$, in the case of
CH$_4$). The atomic correction, $\Delta\Delta\epsilon_C$, corresponds to full
CI, in this case, so we need not discuss it further. It is interesting to note,
however, that it also yields the 'exact' result for C$_2$H$_6$ at large
distances (i.e., the atoms are properly decoupled to separate ground-state
entities), and it deviates by only 1 mH from the full CI value for C$_2$H$_6$
at $f=1$ (using the same basis set). Note that the error of a standard CCSD(T)
calculation is of the same order of magnitude, at that internuclear distance,
and substantially increases for $f > 1$ ($\sim$5 mH for $f$=1.5).

Table 2 gives a compilation of bond increments, $\Delta\epsilon_{CC}$, and atom
increments, $\Delta\Delta\epsilon_C$, for various hydrocarbon molecules, again
for a large range of internuclear distances. It is seen that the
$\Delta\epsilon_{CC}$  are fairly stable in various environments; although
their absolute value changes by more than two orders of magnitude, the maximum
relative change is 3\% between C$_2$H$_6$ and C$_5$H$_{12}$. The
$\Delta\Delta\epsilon_C$  are less transferable: they are invariant, of course,
for $f=100$ as they should, but for $f=1$ the change from four H neighbours (in
CH$_4$) to a purely C-atom neighbourhood (for the central atom in neopentane)
enhances the $\Delta\Delta\epsilon_C$  by nearly a factor of 2. Assuming that
changes in the second-nearest neighbour shell do not appreciably modify
$\Delta\Delta\epsilon_C$ any more, we can make an estimate for the infinite
solid, on the basis of our results. We predict the correlation energy of
diamond, C$_\infty$, for our single-zeta basis set, to be  $\sim 4
\Delta\epsilon_{CC}(C_5H_{12})+2 \Delta\Delta\epsilon_C(C_5H_{12})$ per unit
cell, and obtain -1.8049, -.9921, -.4681, and -.1286 a.u., for $f=100, 2, 1.5,$
and $1$, respectively. 

It is clear that dynamical correlation effects left out in our example, have
significant influence on the properties of diamond, as shown in our previous
work [5,8]. Therefore, we plan to include such effects, in the future, at the
MRCI (or rather MRACPF) level, into our calculations. This would enable a
reliable description of the diamond potential-energy surface up to quite large
internuclear separations. 

\vspace{1cm}

{\section{\bf Summary and conclusions}}

\vspace{0.5cm} 

We have shown that with the help of multicenter scattering matrices the
cumulant wave operator $\mid\Omega)$ can be constructed even when the electron
correlations are strong. The operator $\Omega$ defines the exact ground
state. It follows that the corresponding ground-state energy can be calculated
in form of increments as previously done for weakly correlated electron
systems. The applications described in Section III assumed insulators or
semiconductors, because in that case orthonormal localized SCF orbitals can be
easily constructed. In principle, however, the scattering matrix approach can
be also formulated for nonorthogonal local orbitals.
Within the theoretical framework outlined here accurate ground-state
wavefunction and energy calculations become feasible. It remains a challenging
problem to extend the theory to excited states.

\newpage
\begin{table}
\begin{center}
\caption{Energies of an SCF-like initial wavefunction without/with
subsequent correlation corrections, $\Delta \epsilon_{IJ}$ and
$\Delta\Delta\epsilon_K$ (cf.\ text for definition), as a function of the
bond-length scaling factor $f$, in comparison to full-CI calculations
(FCI). All energies in Hartree.}
\vspace{1cm}
\begin{tabular}{|c|c|c|c|c|} \hline
\multicolumn{5}{|c|}{a) CH$_4$} \\ \hline
$f$ & SCF & + $\sum\Delta \epsilon_{IJ}$ & + $\sum\Delta\Delta\epsilon_K$ &
FCI\\ \hline
100 &  -38.175213 & -39.446762 & -39.696730 & --- \\
2   &  -39.260582 & -39.600272 & -39.760294 & --- \\
1.5 &  -39.727852 & -39.894261 & -39.974412 & --- \\
1   &  -39.990677 & -40.100800 & -40.122505 & --- \\ 
\hline
\hline
\multicolumn{5}{|c|}{b) C$_2$H$_6$} \\ \hline
$f$ & SCF & + $\sum\Delta \epsilon_{IJ}$ & + $\sum\Delta\Delta\epsilon_K$ &
FCI\\ \hline
100 &  -75.661403 & -77.894968 & -78.394909 & -78.394909 \\
2   &  -77.493718 & -78.139308 & -78.485026 & -78.486933 \\
1.5 &  -78.345506 & -78.647679 & -78.833566 & -78.833278 \\
1   &  -78.882457 & -79.059017 & -79.120651 & -79.121427 \\ \hline
\end{tabular}
\end{center}
\end{table}

\begin{table}
\begin{center}
\caption{Correlation corrections for C-C bonds,
$\Delta \epsilon_{CC}$, and atomic re-coupling, $\Delta\Delta\epsilon_C$, for various hydrocarbon molecules,
as a function of the
bond-length scaling factor $f$, cf.\ text. All energies in Hartree.  }
\vspace{1cm}
\begin{tabular}{|c|c|c|c|}
\hline
\multicolumn{4}{|c|}{a) $\Delta\epsilon_{CC}$} \\
\hline
$f$ & C$_2$H$_6$ & C$_3$H$_8$ & C$_5$H$_{12}$ \\
\hline
100 &  -.326241   & -.326241 & -.326241  \\
2   &  -.138377   & -.138113 & -.137592  \\
1.5 &  -.054757   & -.054566 & -.054178  \\
1   &  -.012927   & -.013081 & -.013386  \\
\hline
\hline
\multicolumn{4}{|c|}{b) $\Delta\Delta\epsilon_{C}$} \\
\hline
$f$ & CH$_4$ & C$_2$H$_6$ & C$_5$H$_{12}$ \\
\hline
100 &  -.249968   & -.249970 & -.249970  \\
2   &  -.160022   & -.172859 & -.220869  \\
1.5 &  -.080151   & -.092944 & -.125696  \\
1   &  -.021705   & -.030817 & -.037532  \\
\hline
\end{tabular}
\end{center}
\end{table}


\begin{thebibliography}{99}
\vspace{0.5cm}
\bibitem{ref1} M.C. Gutzwiller, Phys. Rev. Lett. {\bf 10}, 159 (1963)
\bibitem{ref2} J. Friedel: {\sl The Physics of Metals}, ed. by J.H. Ziman
(Cambridge University Press, Cambridge 1969)
\bibitem{ref3} G. Stollhoff and P. Fulde, J.Chem.Phys. {\bf 73}, 4548 (1980)
and earlier references cited therein
\bibitem{ref4} P. Pulay, Chem. Phys. Lett. {\bf 100}, 151 (1983)
\bibitem{ref5} H. Stoll, Phys. Rev. B {\bf 46}, 6700 (1992)
\bibitem{ref6} P. Fulde, {\sl Electron Correlations in Molecules and Solids},
Springer Series in Solid State Sciences, Vol. 100, 3rd edit., (Springer,
Berlin, Heidelberg 1995)
\bibitem{ref7} B. Kiel, G.Stollhoff, C. Weigel, P. Fulde and H. Stoll,
Z.Phys. B {\bf 46},1 (1982) 
\bibitem{ref8} B. Paulus, P. Fulde and H. Stoll, Phys. Rev. B {\bf 51}, 10572
(1995)
\bibitem{ref9} K. Doll, M. Dolg, P. Fulde and H. Stoll, Phys. Rev. B {\bf 55},
10282 (1997)
\bibitem{ref10} S. Kalvoda, M. Dolg, H.-J. Flad, P. Fulde and H. Stoll,
Phys. Rev. B {\bf 57}, 2127 (1998) 
\bibitem{ref11} J. Gr\"afenstein, H. Stoll and P. Fulde, Chem. Phys. Lett. {\bf
215}, 611 (1993)
\bibitem{ref12} H. Stoll, Ann. Phys. {\bf 5}, 355 (1996)
\bibitem{ref13} K. Kladko and P. Fulde, Int. J. Quantum Chem. {\bf 66}, 377
(1998)
\bibitem{ref14} K. Becker and P. Fulde, J. Chem. Phys. {\bf 91}, 4223 (1989)
\bibitem{ref15} R. Kubo, J. Phys. Soc. Jpn. {\bf 17}, 1100 (1962)
\bibitem{ref16} H. D. Ursell, Proc. Cambridge Phys. Soc. {\bf 23}, 685
(1927). J. E. Mayer, J. Chem. Phys. {\bf 5}, 61 (1937)
\bibitem{ref17} L. D. Faddeev, Zh. Eksp. Teor. Fiz. {\bf 39}, 1459 (1960)\\
Engl. transl.: Sov. Phys. - JETP {\bf 12}, 1014 (1961)
\bibitem{ref18} CRYSTAL: a program package described in C. Pisani, R. Dovesi
and C. Roetti {\sl Hartree-Fock Ab Initio Treatment of Crystalline Systems},
Lect. Notes Chem., Vol. 48 (Springer, Berlin, Heidelberg 1988) 
\bibitem{ref19} A. Shukla, M. Dolg, P. Fulde and H. Stoll,
Chem. Phys. Lett. {\bf 262}, 213 (1996)
\bibitem{ref20} C. Hampel and H.-J. Werner, J. Chem. Phys. {\bf 104}, 6286
(1996)
\bibitem{ref21} T.H. Dunning, Jr., J. Chem. Phys. {\bf 90}, 1007 (1989)
\bibitem{ref22} MOLPRO:  an ab initio program package written by H.-J. Werner
and P. J. Knowles, with contributions from J. Alml\"of, R. D. Amos, A. Berning,
M. J. O. Deegan, F. Eckert, S. T. Elbert, C. Hampel, R. Lindh, W. Meyer,
A. Nicklass, K. Peterson, R. Pitzer, A. J. Stone, P. R. Taylor, M. E. Mura,
P. Pulay, M. Sch\"utz, H. Stoll, T. Thorsteinsson, and D. L. Cooper 
\bibitem{ref23} H.-J. Werner and P. J. Knowles, J. Chem. Phys. {\bf 82}, 5053
(1985)
\bibitem{ref24} P. J. Knowles and H.-J. Werner, Chem. Phys. Lett. {\bf 115},
259 (1985)  
\end{thebibliography}
\end{document}